\documentclass[preprint,5p,singlecolumn]{elsarticle}
\journal{Physics Letters B}
\usepackage{color}
\usepackage{amsfonts}
\usepackage{bm}
\usepackage{hyperref}
\usepackage{caption}
\usepackage{subcaption}
\usepackage{amsmath}
\usepackage{slashed} 

\usepackage{array}
\usepackage{longtable}

\begin{document}

\begin{frontmatter}

\title{Triple top signal as a probe of charged Higgs in a 2HDM}

\author{Riley Patrick\fnref{cor1}}
\ead{riley.patrick@adelaide.edu.au}
\author{Pankaj Sharma\fnref{cor2}}
\ead{pankaj.sharma@adelaide.edu.au}
\author{Anthony G. Williams\fnref{cor3}}
\ead{anthony.williams@adelaide.edu.au}
 \fntext[cor1]{ORCID: 0000-0002-8770-0688}
 \fntext[cor2]{ORCID: 0000-0003-1873-1349}
 \fntext[cor3]{ORCID: 0000-0002-1472-1592}
\address{ARC Center of Excellence for Particle Physics at the Terascale, Department of Physics, University of Adelaide, 5005 Adelaide, South Australia}
%\fntext[]{riley.patrick@student.adelaide.edu.au, pankaj.sharma@adelaide.edu.au, anthony.williams@adelaide.edu.au}

\begin{abstract}
Within the framework of the type-II Two Higgs Doublet Model (2HDM-II) we study the production of three top quarks at the Large Hadron Collider (LHC). In the Standard Model the production cross section of three tops is low ($\approx 3$fb), while it is expected to be significant in the 2HDM-II for reasonable choices of the parameters. We study the production of a charged Higgs in association with a top quark, followed by the decays $H^{\pm} \to W^{\pm}A$ and $A \to t\bar{t}$. We undertake a full detector simulation of the signal, and use simple conservative cuts, focussing on the final states that contain three or more leptons, and exactly one same sign di-lepton pair. Finally, we present the exclusion bounds dependent on charged Higgs and pseudoscalar Higgs masses expected in the near future at the 14TeV LHC.
\end{abstract}


\end{frontmatter}

\section{Introduction}
The last missing piece in the standard model (SM) puzzle, the Higgs boson, has been discovered at the Large Hadron Collider (LHC) in its first run. Thereafter various production and decay channels have been studied extensively in order to determine the couplings of the newly discovered boson to various SM particles and the measurements have been found to be close to the SM predictions. Despite this, there is a enough motivation to extend the Higgs sector of the SM in order to understand the underlying mechanism of electroweak symmetry breaking (EWSB). Among several Beyond the SM (BSM) models which include an extended Higgs sector, the two Higgs doublet model (2HDM) is one of the simplest and extensively explored. The particle content of the model is enriched by additional scalars, i.e., two CP even Higgs ($h,~H$), a CP odd Higgs ($A$) and a pair of singly charged Higgs ($H^\pm$). Reviews of the phenomenology of the 2HDM and charged Higgs can be found in ref. \cite{2hdm} and \cite{cHiggs}.

The dominant production of a heavy charged Higgs ($M_{H^\pm}>M_{\rm top}$) at the LHC is in association with a single top quark occurring via $bg\to tH^-$ + c.c. fusion process \cite{Gunion:1986pe}. Charged Higgs decay via bosonic mode i.e., $H^\pm \to W^\pm X$ (where $X\equiv h,H,A$) has received significant attention recently in refs. \cite{bosonic_decays,Arhrib:2016wpw,Li:2016umm,Patrick:2016rtw}. When a heavy neutral Higgs ($H$ or $A$) decays to a top pair, the final state of the process contains a triple top signal and thus would be a unique and interesting probe of charged Higgs at the LHC.

In the SM, the dominant mode of top quark production at the LHC is pair-production, with cross section $\sim 1000$ pb at next-to-leading order (NLO), followed by the single-top quark production with total cross section of 250 pb with the $t$-channel mode having the largest cross section of 150 pb at the NLO.

In addition to single and pair production of top quarks at the LHC, there can also be multi-top quark production in the SM as well, such as three ($3t$) and four top quarks ($4t$). In the SM, the production of an even number of top quarks always occurs via the $gg$ initial state with strong coupling. On the other hand, production of an odd number of top quarks always involves an EW $W^\pm tb$ coupling and often a $b$ quark in the initial state. Thus the cross section for single and three top production is always suppressed with respect to the production of even number of top quarks in the SM. At the LHC with $\sqrt{s}=14$ TeV, the leading order (LO) total cross section for $3t$ production is approximately 1.9 fb while for $4t$ \cite{Barger:2010uw} it is 11 fb, which is almost 6 times the former. In the SM, the $3t$ production occurs via three distinct channels at LO: (a) $pp\to 3t+W^\pm$ at $\mathcal O(\alpha_S^4)$; (b) $pp\to 3t + b$  at $\mathcal O(\alpha_S^2 \alpha_{\rm EW}^2) $; and (c) $pp\to 3t + {\rm jets}$  at $\mathcal O(\alpha_{\rm EW}^4)$. Thus $3t + W^\pm$ has the largest cross section of all $3t$ production modes. 

New physics effects may notably enhance the cross section for $3t$ production over the SM. Thus it could be a sensitive probe of BSM physics. There have been some attempts to investigate new physics in $3t$ production. In ref. \cite{Barger:2010uw}, the authors have studied two BSM models, namely the minimal supersymmetric standard model (MSSM) and the leptophobic $Z^\prime$ model. In the former, the pair production of gluinos and subsequent decays to stops lead to $3t$ production after stop decays via $\tilde{t}\to t \tilde{\chi}^0$. In the latter a $t$-channel exchange of a $Z^\prime$ boson leads to $3t$ production. At a 14 TeV LHC, the production cross section in MSSM is found to be 41 fb, while for the leptophobic $Z^\prime$ model, it is 28 fb. These numbers are significantly larger than the SM cross section of $\sim 2$ fb. In the context of the 2HDM, multi-top production including both $3t$ and $4t$ production has been studied at the LHC and for the international linear collider (ILC) in ref. \cite{Kanemura:2015nza}. This study utilizes the subdominant charged Higgs production processes which are associated $H^\pm H$ and $H^\pm A$ production followed by $H^\pm \to tb$ and $A/H\to t\bar t$ decays. 
% However only subdominant modes of charged Higgs production i.e. $pp\to H^\pm H (A)$ where $H$ and $A$ denote the heavy CP even Higgs and CP odd Higgs respectively were taken into account. Then further decays of $H^\pm\to t b$ and $H (A)\to t\bar t$ were considered to get triple-top production.

In this letter, we will focus primarily on the triple top production $pp\to 3t+X$, facilitated by a charged Higgs and a pseudoscalar in a two Higgs doublet model (2HDM). We will demonstrate that $3t$ production can be an alternative probe of the charged Higgs at the LHC, especially for the scenario where both the charged Higgs and pseudoscalar are much heavier than top quarks.  In our analysis, we make use of the dominant production mode of a heavy charged Higgs at the LHC, $pp\to t H^-$ followed by the decay of charged Higgs via the bosonic mode, $H^\pm \to W^\pm A$ and pseudoscalar Higgs via the $A\to t\bar t$ mode. Thus it leads to three top quarks in the final state. We perform a realistic simulation of the triple top signal, including detector effects, and apply a set of kinematical cuts to suppress the backgrounds. We present exclusion/discovery bounds after including the effects of all the irreducible and reducible backgrounds in the plane of charged Higgs mass and pseudoscalar mass for a 14 TeV LHC with 30 fb$^{-1}$ of integrated luminosity.

The plan of the paper is following. The next section discusses the production cross section and decay branching ratio of the charged Higgs and triple top signal. Section 3 discusses the different signals and their corresponding backgrounds. In section 4, we present our results after a full simulation and analysis of the events. Finally, we conclude and summarize in section 5.

\section{Production and Decay}
\begin{figure}[h!]
	\includegraphics[scale=0.2]{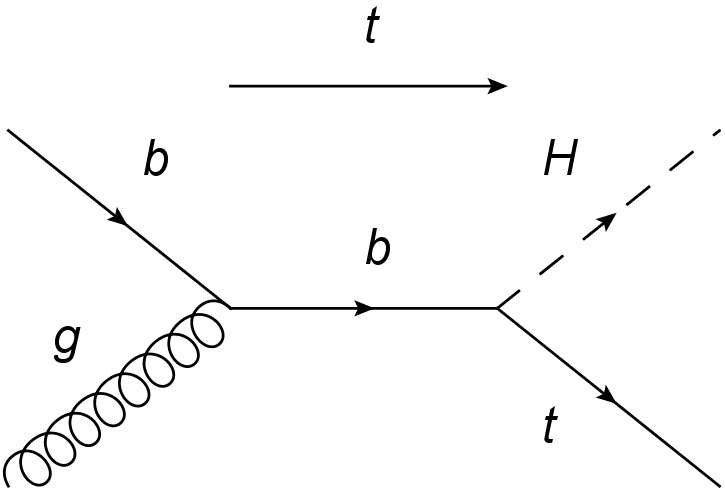}\hspace{5mm}
	\includegraphics[scale=0.2]{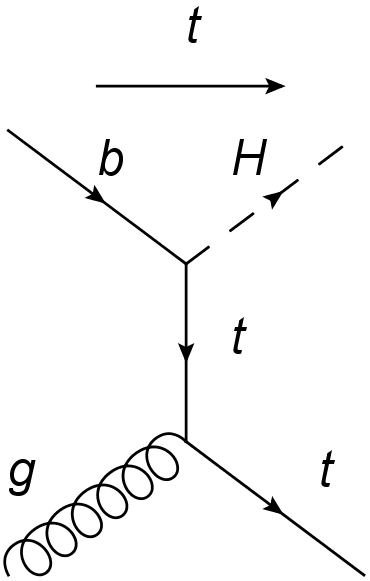}
	\caption{\label{fig:feyn} The leading order production mechanisms for $pp\to tH^{\pm}$}
\end{figure}

\begin{figure}[t!]
\centering
	\includegraphics[scale=1.]{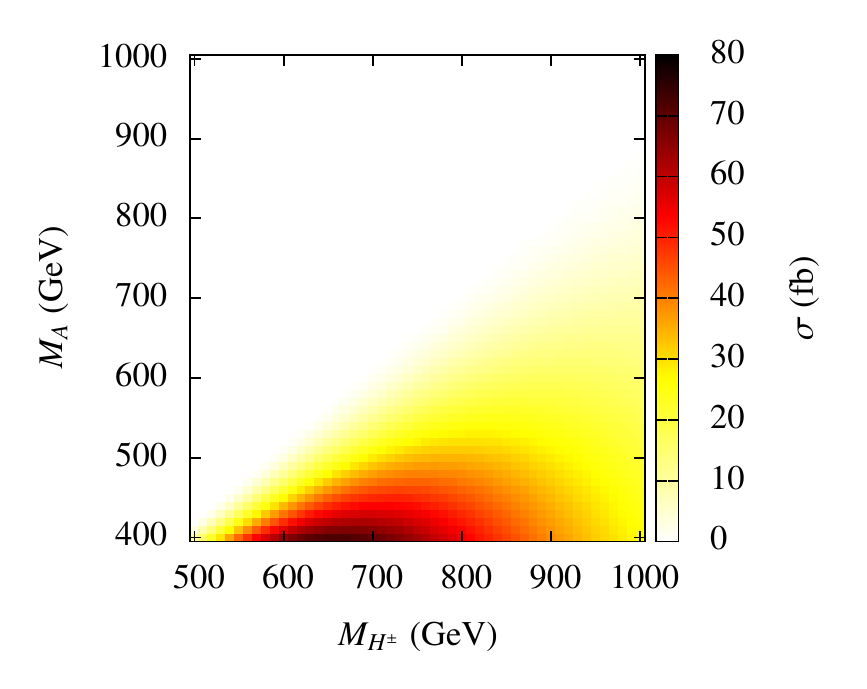}
	\caption{\label{fig:cs} Cross section for triple top signal ($tt\bar t W^\pm$) obtained by evaluating $\sigma(pp\to tH^\pm)\times$Br($H^{\pm}\rightarrow WA$)$\times$Br($A\rightarrow t\bar{t}$) in the plane of charged Higgs mass and pseudo scalar mass.}
\end{figure}

The process considered in this analysis is a charged Higgs production in association with a single top quark, the leading order Feynman diagrams can be seen in Fig.~\ref{fig:feyn}. 
Following the production we consider the decay $H^{\pm} \rightarrow W^{\pm}A$ and $A\rightarrow t\bar{t}$ and thus leading to three top quarks in the final state. In Fig.~\ref{fig:cs} we present the production cross section of the triple top signal in type-II 2HDM obtained by multiplying the cross section of the process $pp\to tH^-$ with the branching ratios of the decays described above i.e., $H^\pm\to W^\pm A$ and $A\to t\bar t$ in the plane of the charged Higgs mass and pseudoscalar mass. We use the Two Higgs Doublet Model Calculator (\texttt{2HDMC}~\cite{Eriksson:2009ws}) to obtain the corresponding branching ratios for each point in the parameter space. In order to evaluate the cross section and BRs, we consider $\tan\beta=1$ and $\sin(\beta-\alpha)=1$ throughout the analysis. 

% Thus, we are demanding that the pseudo-scalar have mass $\geq 350$ GeV and the charged Higgs have mass $\geq 420$ GeV, which fits nicely with the $b\rightarrow s\gamma$ constraints for the type-II 2HDM which renders $M_{H^{\pm}}\geq 580$ GeV. We also consider the scenario in which the CP-even heavy higgs, $H$, has mass $M_H\approx M_{H^{\pm}}$, however if this were not the case one could simply adjust the branching ratios of the charged Higgs decay, and then consider the additional channel of $H\rightarrow t\bar{t}$ and apply the same analysis.

The mass splitting between the $H^\pm$ and $A$ should be greater than 80 GeV so as to achieve a large BR to the $W^\pm A$ mode. The charged Higgs mass $M_{H^\pm}$ and pseudoscalar mass $M_A$ are varied in the ranges (500 GeV - 1 TeV) and (400 GeV - 1 TeV) respectively. Also, in order to open the $A\to t\bar t$ decay channel, the pseudoscalar mass should be larger than $2\rm M_{top} \sim 350$ GeV. For such a pseudoscalar the dominant decay mode would be to a top pair as the coupling is proportional to mass of the top quark. The other decay $A\to Zh$ is suppressed by $\sin(\beta-\alpha)$ since the current LHC scenario prefers alignment scenario with $\sin(\beta-\alpha)=1$. Therefore in the scenario where both $M_{H^\pm}$ and $M_A$ are heavier than $2M_{\rm top}$, the triple-top production is the only possible signal to probe a charged Higgs at the LHC. 

We see from Fig. \ref{fig:cs} that large cross section for triple top production in a 2HDM is obtained when $M_{H^\pm}$ is in the range (550 GeV - 750 GeV) and $M_A$ in the range (400 GeV - 500 GeV), where the cross section is found to be in the range (50 fb - 80 fb). This is significantly larger than the SM cross section of 2 fb for triple top quarks at the 14 TeV LHC. Thus it is expected that the search for the triple top signal in the current and future runs at the LHC would significantly enhance the search prospects for a charged Higgs. If no such signal is found, it would enable a stringent bound to be set in the $M_{H^\pm}-M_A$ plane in 2HDM. 

\begin{figure}[t!]
	\includegraphics[scale=0.5]{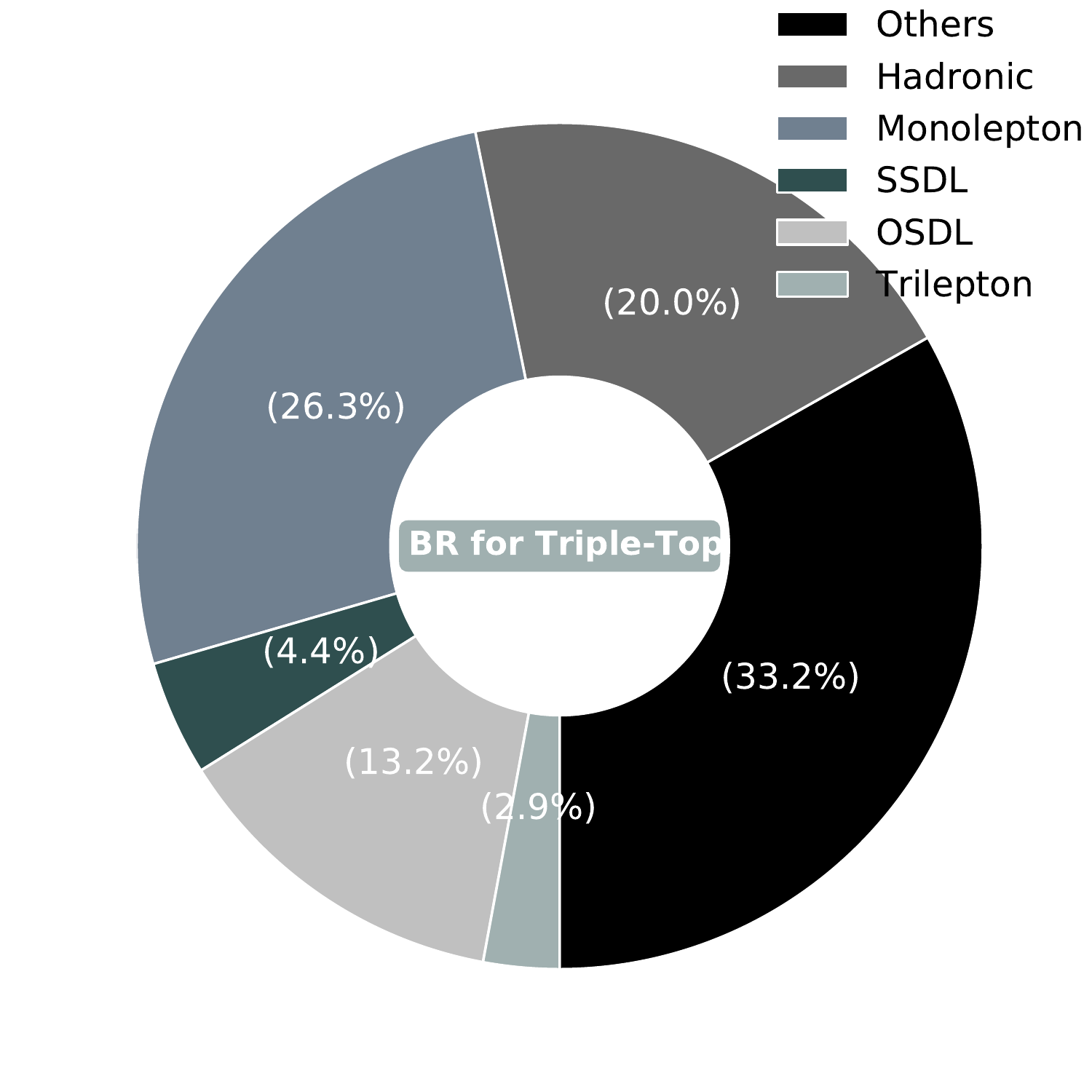}\hspace{5mm}
	\caption{The branching ratio associated with each possible final state for $tt\bar{t}+W^-$ and $t\bar{t}\bar{t}+W^+$ decays. \label{fig:BR}}
\end{figure}

\section{Signal and SM Backgrounds}
\subsection{Signal}
The triple-top event $tt\bar t W^\pm$ decays into a large number of final state particles, $\sim\mathcal O(10)$. Various decay modes and their corresponding branching ratios have been presented in Fig.\ref{fig:BR}. Fully hadronic decays of the triple top event which has 20 \% of BR leads to a very high number of jets in final state, i.e., 8 light jets and 3 $b$ jets. The monolepton signal includes one lepton (either $e^\pm$ or $\mu^\pm$, $\tau^\pm$ is not included here) associated with 9 jets and has fairly large BR of 26.3\%. The dilepton signal can be classified into opposite sign dilepton (OSDL), which has 14.2\% BR, and same-sign dilepton (SSDL), which has quite small 4.4\% of BR. Finally the trilepton signal has only a 2.9\% of BR. 

In this analysis we focus on the multileptonic signals, namely SSDL and trilepton, which despite having low branching ratios are cleaner at the LHC and have backgrounds which are more efficiently manageable. The  SSDL signal is accompanied by 7 additional jets while the trilepton signal arises along with 5 jets. In both cases, three of the jets are $b$ jets.

\subsection{Backgrounds}
The irreducible background to the final state being considered is the SM production of three tops in association with a $W^\pm$ boson, which has a total production cross section of $1.37$ fb. However, it is reasonable to expect any three top signal to behave as a background to the signal under the right circumstances (IS/FS radiation, jet mis-tagging etc.). However, the combined cross section of all three top production as the LHC is only $1.9$ fb.

As well as this in the circumstance that a b-jet from the final state is lost from the detector, a four top process will successfully mimic our signal. This is concerning given the cross section of four top production at the LHC is much higher at $11$ fb, though cuts on pseudo-rapidity and jet multiplicity should almost certainly remove most of these processes.

Other backgrounds come from various top pair production associated with one heavy SM particle and light jets of which at least one must be $b$ jets, for instance, $t\bar t W^\pm$, $t\bar t h$ and $t\bar t Z$ processes. Processes such as $t\bar t ~b ~nj$ may contribute to both trilepton and SSDL signals when one of the light jets are faked as a lepton. $Q-$flip backgrounds occur when a real OSDL signal arising from some underlying process is misidentified as SSDL pair at the detector. 

\section{Signal and Background Efficiency}
The accumulative efficiency associated with detector effects, jet finding/tagging and kinematic cuts is important to get a full understanding of signal significance. To achieve this we generate a million events using {\tt Madgraph} \cite{madgraph}, which are passed to {\tt Pythia} \cite{pythia8} for parton showers and hadronization and finally to {\tt Delphes} \cite{delphes} for realistic detector effects. For detector acceptance, we apply the following cuts: (i) all leptons must have transverse momentum $p_T$ larger than 20 GeV and be within pseudorapidity $|\eta|<2.5$; (ii) all the jets must have $p_T>25$ GeV and $|\eta|<2.5$; (iii) all the objects must be well separated from with each other with $\Delta R_{ij}>0.4$ where $\Delta R=\sqrt{(\Delta \phi)^2+(\Delta \eta)^2}$. All the jets are clustered using an anti-$k_T$ jet clustering algorithm with jet radius $\Delta R=0.4$.

We then choose two final states to study, the trilepton and SSDL final states. In ref.\cite{Alvarez:2016nrz}, authors have studied four-top production and analysed the event in SSDL and trilepton signals at the 14 TeV LHC. The corresponding backgrounds to these signals are also modelled in great detail. We adopt the search strategy for selection of signal and background events presented in their analysis. In the tri-lepton case we demand that the event contains more than 5 light jets ($n_{\text{jet}}>5$) and three or more b-tagged jets ($n_{\text{bjet}}\geq 3$). We also require that all same flavour opposite sign leptons invariant mass does not fall within 25 GeV of the $Z$-boson mass which serves to significantly cut into the SM background. In the SSDL case we use $n_{\text{jet}}>6$ and $n_{\text{bjet}}\geq 3$. 

The final accumulative signal efficiency for the trilepton and SSDL final states after the above selection requirements are $49.9\%$ and $77.2\%$ respectively. The difference in these efficiencies can be attributed to the extra $Z$-boson mass window cut placed on the trilepton state.

As mentioned above, we adopt the same set of cuts employed in ref.~\cite{Alvarez:2016nrz} on the signal events. Thus in this analysis, we use the same efficiencies for the backgrounds presented in their study. The total background cross section for the trilepton signal is $60.67$ fb, while for the SSDL signal it is $122.73$ fb. As well as this the efficiencies associated with each background can be seen in Table~\ref{tab:ss_ll_eff} and Table~\ref{tab:lll_eff}. 

\begin{table}[t!]
\begin{center}
\begin{tabular}{ || c | c | c ||}\hline 
Backgrounds & Cross Section (fb) & Efficiency \\\hline
ttW & 31.34 & $6.46\times 10^{-3}$ \\
ttZ & 48.47 & $2.21\times 10^{-3}$ \\
tth & 7.25 & $2.53\times 10^{-3}$ \\
Fakes & 16.57 & $3.48\times 10^{-3}$ \\
Q-flip & 11.37 & $1.01\times 10^{-2}$ \\
Other & 7.73 & $1.90\times 10^{-3}$ \\\hline
\end{tabular}
\caption{The background cross sections and corresponding efficiencies for the SSDL signal at the 14 TeV LHC. \label{tab:ss_ll_eff}}
\end{center}
\end{table}
\begin{table}[h!]
\begin{center}
\begin{tabular}{ || c | c | c ||}\hline 
Backgrounds & Cross Section (fb) & Efficiency \\\hline
ttW & 1.65 & $9.90\times 10^{-3}$ \\
ttZ & 48.47 & $5.50\times 10^{-4}$ \\
tth & 2.4 & $2.50\times 10^{-3}$ \\
Fakes & 1.13 & $5.01\times 10^{-3}$ \\
Other & 7.02 & $4.70\times 10^{-4}$ \\\hline
\end{tabular}
\caption{The background cross sections and corresponding efficiencies for the trilepton signal at the 14 TeV LHC. \label{tab:lll_eff}}
\end{center}
\end{table}

\begin{figure}[h!]
\centering
\includegraphics[scale=0.475]{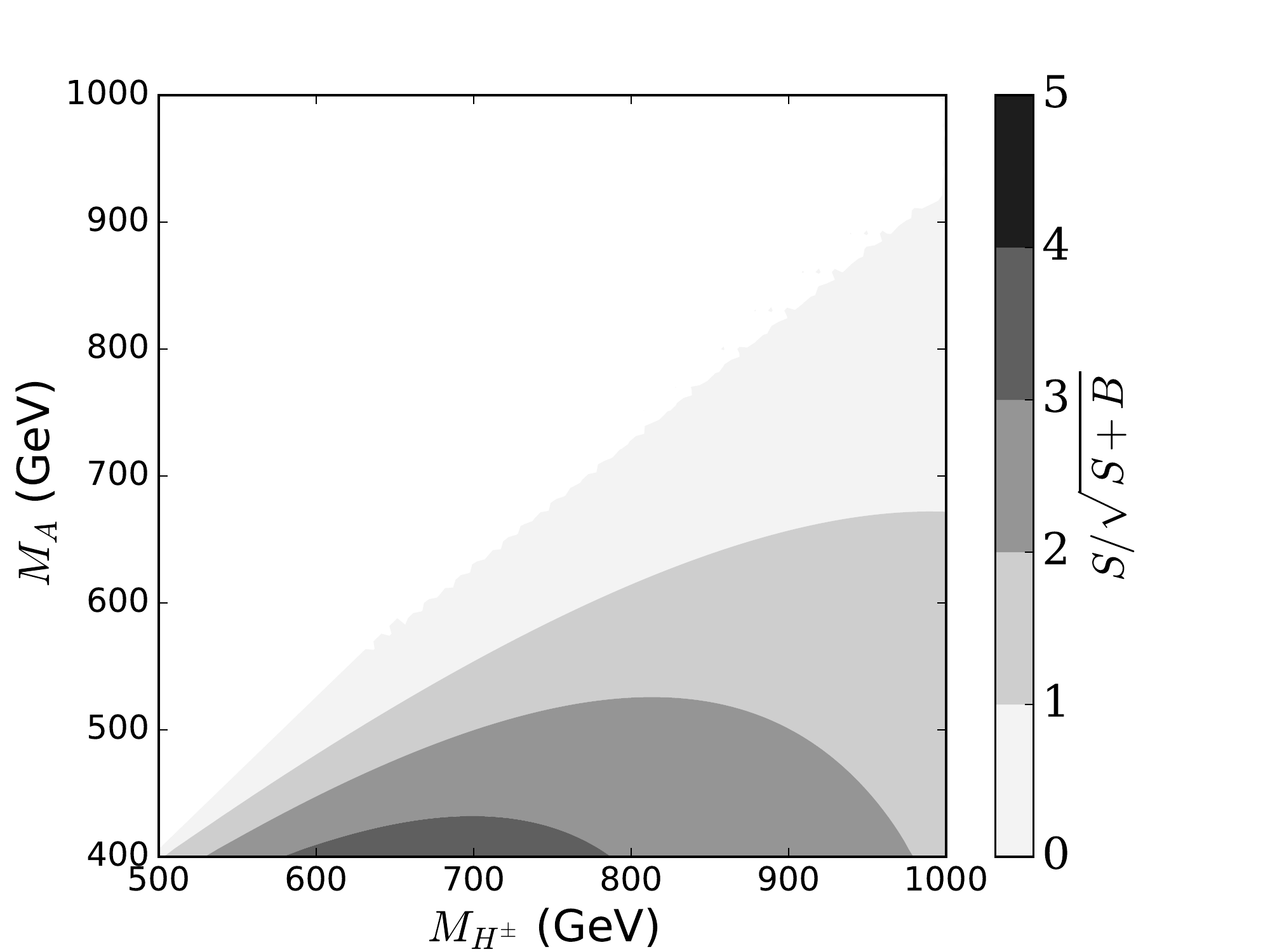}
\includegraphics[scale=0.475]{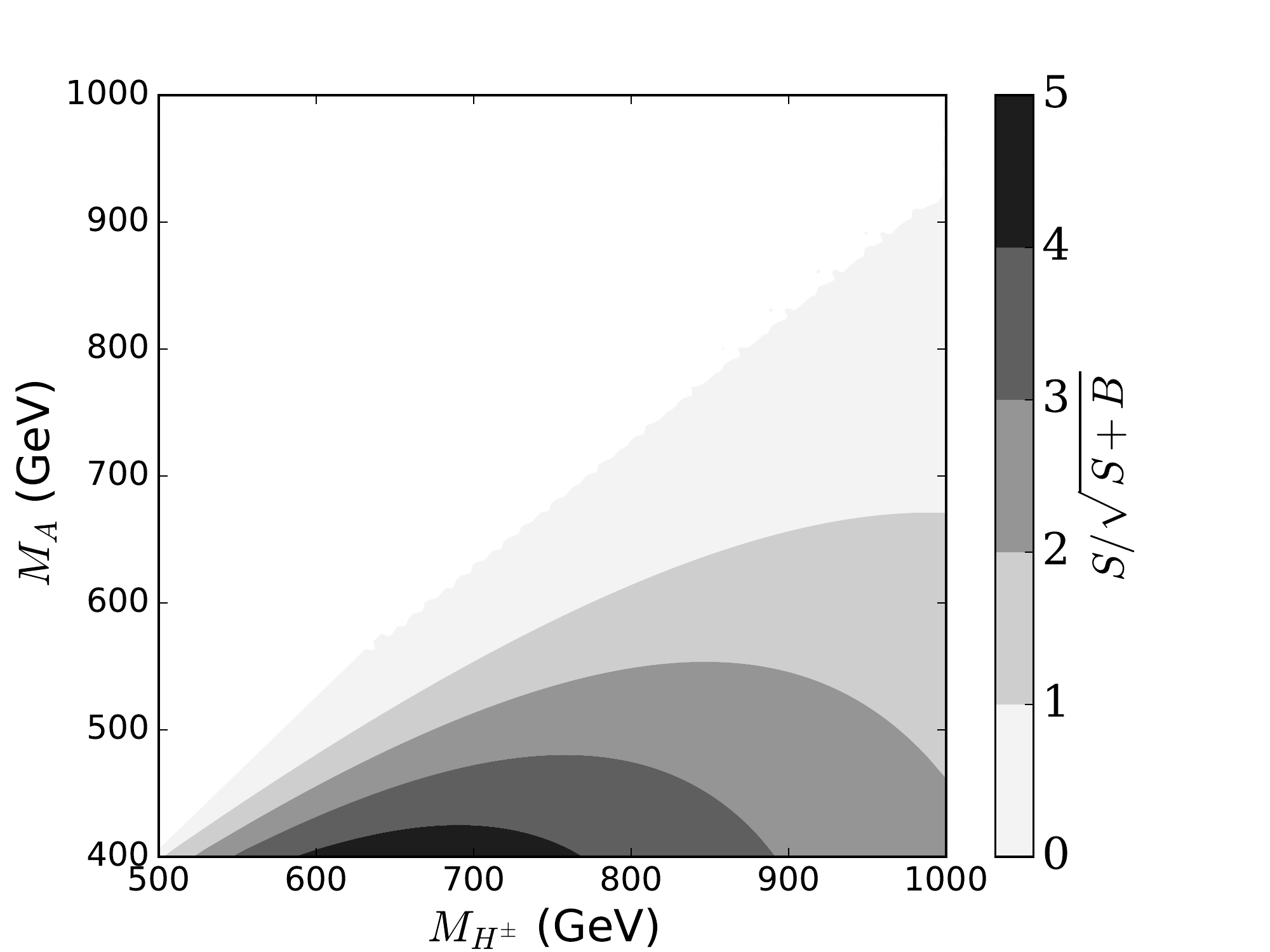}
\caption{\label{fig:exclusion} Significance for charged Higgs exclusion/discovery for the trilepton (top) and SSDL signal (bottom) with 30 fb$^{-1}$ of integrated luminosity at the 14 TeV LHC.}
\end{figure}

% \subsection{Exclusion Bounds}
Using the efficiencies for signals and background obtained above we now estimate the signal significance which is defined by the ratio $S/\sqrt{S+B}$ where $S$ and $B$ are the number of signal and background events respectively. We present in the Fig.~\ref{fig:exclusion} the signal significance using trilepton (top) and SSDL (bottom) signals in the plane of mass of the charged Higgs and mass of the pseudoscalar for the 14 TeV LHC with 30 fb$^{-1}$ of integrated luminosity. As expected, we find the SSDL signals to be more constraining than the trilepton signal due to the larger cross section and smaller backgrounds as compared to the latter. We conclude that with the early data to be collected in the 14 TeV LHC, the SSDL and trilepton signal in the triple-top production can exclude the charged Higgs upto 1 TeV if the mass splitting between $H^\pm$ and the pseudoscalar is within the range (100 GeV - 300 GeV). 

\section{Conclusion}

In this letter we have assessed the discovery/exclusion potential of a heavy charged Higgs boson in the type-II Two Higgs Doublet Model (2HDM) at the large hadron collider (LHC). We explored the case when a charged Higgs with mass $\geq 450$ GeV is produced in association with a top quark, followed by the decay $H^{\pm}\to AW^{\pm}$ where $A$ is the pseudo-scalar Higgs. This pseudoscalar Higgs was chosen to be $\geq 350$ GeV so as to allow its decay to two top quarks, resulting in the state $t\bar{t}\bar{t}W^{+}$ or $tt\bar{t}W^{-}$. This is an exotic state that is extremely rare in the standard model (SM) (triple-top production in the SM has a cross section of less than $3$ fb) and as such provides an interesting window into the search for charged Higgs in the 2HDM. 

We further focused on the final state that included trilepton and also the final state with a same sign dilepton (SSDL) pair. The discovery/exclusion potential for these signals has been presented in Fig~\ref{fig:exclusion}. It can be seen that with current LHC data, with approximately $30$ fb of luminosity, we should be able to make strong statements regarding the existence of these states in almost all parts of the charged Higgs/pseudo-scalar mass plane for our chosen parameters. 

Further directions for this work revolve predominantly around cut optimization. The above analysis used extremely conservative cuts and achieved great exclusion potential but the use of more advanced techniques such as Boosted Decision Tree or Artificial Neural Nets as seen in Refs.~\cite{Li:2016umm}~\cite{Patrick:2016rtw} should significantly improve the exclusion potential but would require the full modelling of the SM backgrounds at the detector level. This was not undertaken in this study. 

In addition, the remainder of the final states - opposite sign dilepton pair, monolepton and fully hadronic - could provide some level of exclusion potential but these are expected to be lower than the channels presented in this study. This is because while the branching ratios of these final states are higher, the SM backgrounds are much higher and may drown out the signal.

Finally, a study of the interference effects between the signal and irreducible SM backgrounds would be of great interest. It is expected that the interference terms of the full scattering amplitude would produce a peak dip structure in the distributions of the variables of the events. This may be exploited to further improve the discovery/exclusion potential or it may obscure the signal.

\section*{Acknowledgements}
This work is supported by the University of Adelaide and the Australian Research Council through the ARC Center of Excellence for Particle Physics (CoEPP) at the Terascale (grant no. \ CE110001004).

\end{document}